\documentclass[notitlepage,twocolumn,prb,aps,showpacs,10pt,superscriptaddress]{revtex4-1}

\usepackage{amsmath}
\usepackage{graphicx}
\usepackage{xcolor}
\usepackage{hyperref}   							
\hypersetup{colorlinks=true,allcolors=blue}

\newcommand{\mbf}[1]{\mathbf{#1}}
\renewcommand{\t}[1]{\textrm{#1}}

\renewcommand{\k}{\mbf{k}}
\newcommand{\R}{\mbf{R}}
\newcommand{\ddt}{\frac{\partial}{\partial t}}

\newcommand{\nn}{\nonumber\\}

\begin{document}
\title{Nonexponential spin decay in a quantum kinetic description 
of the D'yakonov-Perel' mechanism mediated by impurity scattering}
\author{M. Cosacchi}
\affiliation{Theoretische Physik III, Universit{\"a}t Bayreuth, 95440 Bayreuth, Germany}
\author{M. Cygorek}
\affiliation{Theoretische Physik III, Universit{\"a}t Bayreuth, 95440 Bayreuth, Germany}
\author{F. Ungar}
\affiliation{Theoretische Physik III, Universit{\"a}t Bayreuth, 95440 Bayreuth, Germany}
\author{V. M. Axt}
\affiliation{Theoretische Physik III, Universit{\"a}t Bayreuth, 95440 Bayreuth, Germany}
\begin{abstract}
The electron spin dynamics in an optically excited narrow quantum well is studied, where 
the electron spins precess in a $\mathbf{k}$-dependent magnetic field while 
the electrons scatter at localized impurities. 
For the resulting spin decay, which is commonly known as the D'yakonov-Perel' mechansim, 
analytical expressions in the strong- and weak-scattering limits are available.
It is found by the numerical solution of quantum kinetic equations in a broad
range of parameters that, 
in situations that are typically relevant for ultrafast optical experiments, 
the dynamics of the total spin polarization significantly deviates 
from the pertinent analytical results. This is attributed to the broad 
spectral width of the optically excited spin-polarized electron distribution,
which gives rise to a spin dephasing due to inhomogeneous broadening.
Furthermore, it is found that the decay of the spin polarization need not be
exponential. The notion of a spin decay time becomes ambiguous and different definitions of spin decay times can lead to different outcomes.
The long-term dynamics of the decay of the spin polarization is even dominated by
an algebraically decaying component.
These findings highlight the importance of the effects of the broad spectral
distribution of optically excited carriers in ultrashort magneto-optical experiments.
\end{abstract}
\maketitle

\section{Introduction}
In the field of spintronics 
\cite{Zutic,Spintronics,spintronics_dietl,spintronics_ohno}, the spin dynamics in 
semiconductors has attracted a lot of interest in the last decades since the spin 
dephasing times in semiconductors can be orders of magnitude larger than, e.g., in 
metals\cite{Awschalom07,Ohno_GaAs110}. There are two different approaches for 
studying the spin dynamics in semiconductors which have to be distinguished:
transport 
\cite{Zutic,DattaDas,DAmico_CoulombDrag,FabianZuticReview_ActaPhysSlov,SpinHallExp,SpinInjAlGaAs,TransportAppelbaum,Crooker_ImagTransport,Fiederling} 
and optical experiments
\cite{optorient,Awschalom_RSA,Korn_2DEGDephasingGaAs,Korn2DHG,MarieHoleG_GaAs,Crooker97}. 
The former uses the fact that the injection of carriers into the semiconductor and 
the transmission from the semiconductor into another material can be strongly spin-
dependent. Therefore, the resistivity of devices consisting of a semiconductor 
material between a spin injector and a spin filter is strongly affected by the 
dynamics of the carrier spins in the semiconductor \cite{Zutic,DattaDas}, which paves 
the way for the development of spin transistors \cite{DattaDas,SchliemannSFET}. 
Optical experiments, on the other hand, allow a very direct control and readout of 
the carrier spins in a semiconductor structure via the spin selection rules and the 
magneto-optical Faraday and Kerr effect, e.g., in optical pump-probe measurements
\cite{Awschalom_RSA,Korn_2DEGDephasingGaAs,Korn2DHG,MarieHoleG_GaAs,Crooker97}.

The literature on spin dynamics in semiconductors \cite{WuReview} 
typically lists the D'yakonov-Perel'\cite{DP} (DP), the 
Elliot-Yafet \cite{EY1,EY2,EY_Baral} (EY) and the Bir-Aronov-Pikus\cite{BAP} (BAP) mechanisms
as the main sources for the decay of a non-equilibrium carrier spin 
polarization. 
The BAP mechanism is due to the interaction between electron and hole spins and it is 
therefore mostly relevant for the electron spin dynamics in p-doped semiconductors 
\cite{WuBAP_pGaAs}. Both the DP and the EY mechanisms originate from spin-orbit 
interaction (SOI) and the mixing of conduction and valence bands for non-zero wave 
vectors $\k$ according to $\k.\mbf p$-theory. The EY mechanism is based on the fact 
that, due to the band mixing for finite wave vectors $\k$, the energy eigenstates of 
the quasi-free carriers are no longer eigenstates of the spin operator. Thus, 
scattering of electrons leading to a change of the wave vector $\k$ also leads to a 
change in the average electron spin. Another effect resulting from the mixing of 
valence and conduction bands is that a block-diagonalization which eliminates the 
coupling of different bands renormalizes the crystal Hamiltonian, so that the 
conduction band block after block-diagonalization acquires an additional term that 
can be written in the form of a Zeeman energy with a $\k$-dependent effective 
magnetic field. The DP mechanism describes the combined effect of the 
dephasing of electron spins caused by the precession in the strongly anisotropic $\k
$-dependent field and momentum scattering of the electrons at, e.g., impurities, 
other carriers or phonons. 
In the strong-scattering limit, D'yakonov and Perel' have derived \cite{DP} their
well-known result that the spin decay rate is inversely proportional to
the momentum scattering rate. Also in the weak-scattering limit, an analytical
relation between spin decay and momentum scattering can be obtained. \cite{Gridnev2001,WuReview}
However, there, the spin decay rate is proportional to the momentum scattering rate.

Another mechanism leading to a decay of the total electron spin polarization 
has been pointed out by Wu and Ning \cite{NingDP,WuNingEuroPhys}:
the spin precession frequency given by the magnitude of the effective field 
depends on the modulus $|\k|$ of the wave vector. Thus, when electrons with 
different kinetic energies take part in the spin dynamics, the presence of
different precession frequencies gives rise to a dephasing of spins even in the absence of 
momentum scattering, where the conventional DP mechanism predicts no spin decay. This 
effect is referred to as inhomogeneous broadening by Wu \textit{et. al.} in
Refs.~\onlinecite{NingDP,WuNingEuroPhys} and is characterized by an algebraic spin decay
$\propto \frac 1t$ for long times \cite{WuReview}. In order to distinguish this mechanism
from other effects typically associated with the term inhomogeneous broadening in 
the context of optical experiments on semiconductors, such as the 
linewidth broadening of excitons caused by spatial fluctuations of the environment
\cite{ExcLinewidthInhomBroad},
in the following we use the term \textit{dispersion-induced isotropic inhomogeneous broadening} (DIIB) for the spin dephasing induced by the $|\k|$-dependence of the
effective magnetic field.

Numerous works in the literature have addressed the question of how 
the spin dynamics in semiconductors is affected by the DP, the EY and the BAP
mechanisms.\cite{WuReview,Lu2006,Leyland2007,Brand2002,Pershin2007,Chazalviel1975,Jiang2009,Cheng2010,
Maialle1996,Wu2000} In particular, 
Wu \emph{et al.} \cite{WuReview} have reviewed the contribution of 
different spin dephasing and momentum scattering mechanisms to the total spin decay 
time using kinetic spin Bloch equations (KSBE).
Most works focus on calculating spin transfer rates and their dependencies on 
certain parameters such as carrier concentration, temperature, external magnetic 
field, and so on.\cite{SpringerDyakonov2008,Kavokin2008} However, rates are only well-defined if the spin decay is 
approximately exponential, which is \emph{a priori} not clear. 
If the spin decay is nonexponential, the concept of a spin decay rate becomes ill-defined.

In this article, we investigate the time evolution of the electron spin in an 
Al$_{x}$Ga$_{1-x}$As semiconductor quantum well (QW) after optical 
excitation with circularly polarized light. We consider the precession of 
electron spins in a $\k$-dependent Dresselhaus or Rashba field and scattering of 
electrons at Al impurities. This is a situation which is conventionally described by 
the DP mechanism. Here, however, we use a microscopic quantum kinetic density matrix 
theory, which goes beyond the conventional DP picture in several aspects: 
First, by resolving the 
$\k$-space we explicitly consider an ensemble of electrons, whereas the standard DP 
theory \cite{DP} describes a stochastic motion of a single electron. Thus, our 
theory includes the spin dephasing due to DIIB 
\cite{NingDP,WuNingEuroPhys}. Second, we do not \emph{a priori} postulate the existence 
of a well-defined spin decay rate, i.e., we calculate the time evolution of the total 
electron spin explicitly and do not assume that it is exponential. This allows us to 
study the spin dynamics even when the notion of a spin decay rate is questionable. 
Finally, the quantum kinetic description goes beyond perturbation theory in 
the carrier-impurity interaction, the Markov approximation and the single-particle 
picture as it includes explicitly correlations between carriers and impurities that 
are built up during the scattering. Our theory is applicable not only in the limiting 
cases of weak and strong scattering, but also in the intermediate regime and, 
therefore, allows us to study the range of validity of the results in the limiting 
cases.

We find that the time evolution of the total spin polarization after optical excitation
can have different shapes, ranging between an exponentially damped oscillation to
a Gaussian-like monotonic decay. There are also situations where the spin decays 
highly nonexponential, has a minimum, and shows a slow decay at large times. 
Furthermore, the long-term dynamics can be
dominated by an algebraic decay.
In particular, we find that the broad width of the optically induced electron distribution
has very important effects on the spin dynamics, highlighting the importance 
of DIIB for ultrafast optical experiments.
Furthermore, we find that because of the nonexponential nature of the
spin dynamics different definitions of a characteristic spin decay time
can give quantitatively and qualitatively different results.
This shows that the concept of a spin decay time has to be treated with care
when discussing results of ultrafast optical experiments.

The paper is structured as follows: first, we set up a quantum kinetic theory
for the dynamics of the electron density matrix as well as electron-impurity
correlations. Subsequently, we derive the Markov limit of the quantum kinetic equations 
of motion and discuss theoretically certain known limiting cases. 
Then, we present numerical results of Markovian and quantum kinetic calculations 
for an optically excited Al$_x$Ga$_{1-x}$As quantum well with Dresselhaus and
Rashba spin-orbit field. Finally, we summarize the results.

\section{Theory}
We study the spin dynamics in a semiconductor quantum well after optical excitation
which can be experimentally addressed by optical pump-probe measurements such as in 
time-resolved Kerr rotation experiments \cite{Crooker97}.
More specifically, we consider a D'yakonov-Perel'-type system \cite{DP} where
the optically induced electron spins precess in a $\k$-dependent effective
magnetic field like the Dresselhaus \cite{Dresselhaus} or Rashba \cite{Rashba} 
field and the carriers are subject to momentum scattering at localized 
impurities. Depending on the sample and the excitation conditions, there are also
situations in which other momentum relaxation mechanisms are dominant, such as
phonon scattering or carrier-carrier scattering \cite{WuBAP_pGaAs,WuReview}. 

Here, however, we consider a
narrow Al$_{x}$Ga$_{1-x}$As quantum well with Al$_{y}$Ga$_{1-y}$As barriers,
where $y>x$ to ensure the confinement of carriers in the well. We focus on 
a situation where the Al concentration $x$ in the quantum well is not too small, the
temperature of the sample is low enough to suppress phonon scattering and the
intrinsic sample is excited with low or moderate intensity so that carrier-carrier
interactions are of minor importance. 
Then, the momentum scattering at the Al impurities 
dominates and other momentum scattering mechanisms are negligible.
Furthermore, we assume that the spins of the optically induced holes dephase fast
due to the strong spin-orbit interaction in the valence band and we are 
only interested in the dynamics of the conduction band electron spins.
Moreover, the quantum well is assumed to be narrow enough so that only the
lowest confinement state has to be considered and the relevant electronic states
can be described by plane waves with two-dimensional in-plane wave vectors $\k$.

\subsection{Hamiltonian}
The Hamiltonian for conduction band electrons in a
narrow Al$_{x}$Ga$_{1-x}$As quantum well is
\begin{align}
\label{eq:H}
\hat{H}=\hat{H}_0+\hat{H}_{SO}+\hat{H}_{\textrm{Imp}}\, ,
\end{align}
where
\begin{align}
\label{eq:H0}
\hat{H}_0&=\sum_{\sigma\k}\hbar\omega_\k \hat{c}_{\sigma\k}^\dagger\hat{c}_{\sigma\k}^{}
\end{align}
describes the spin-independent part of the band structure, which we assume to be parabolic according to $\omega_\k=\frac{\hbar\k^2}{2m^*}$
with the two-dimensional in-plane wave vector $\k$ and in-plane effective mass $m^*$. 
The symbol $\sigma\in\{\uparrow,\downarrow\}$ denotes the spin index and distinguishes the
two conduction subbands.
Finally, $\hat{c}_{\sigma\k}^\dagger$ and $\hat{c}_{\sigma\k}^{}$ are the electron creation and annihilation operators, respectively. 

The spin-orbit interaction (SOI) is described by the Hamiltonian
\begin{align}
\label{eq:HSO}
\hat{H}_{SO}&=\sum_{\sigma\sigma'\k}\hbar\mathbf{\Omega}_{\k}
\cdot\mathbf{s}_{\sigma\sigma'}\hat{c}_{\sigma\k}^\dagger\hat{c}_{\sigma'\k}^{}\, .
\end{align}
Here, $\mathbf{s}_{\sigma\sigma'}=\frac{1}{2}\mathbf{\sigma}_{\sigma\sigma'}$, where $\boldsymbol{\sigma}$ denotes the vector of Pauli matrices
and $\mathbf{\Omega}_{\k}$ is the $\k$-dependent effective magnetic field
arising from bulk (BIA) or structure inversion asymmetries (SIA) that result 
in Dresselhaus \cite{Dresselhaus} and Rashba \cite{Rashba} contributions 
to the effective magnetic field of the form 
\cite{Winkler,Ganichev14}
\begin{subequations}
\label{eq:OmegaRashbaDresselhaus}
\begin{align}
\mathbf{\Omega}_{\k_1}=\mathbf{\Omega}_{\k_1}^{\textrm{Rashba}}
+\mathbf{\Omega}_{\k_1}^{\textrm{Dresselhaus}}
\end{align}
\begin{align}
\label{eq:OmegaRashba}
\mathbf{\Omega}_{\k_1}^{\textrm{Rashba}}=2\frac{\alpha_R}{\hbar}\left(\begin{array}{c} k_y \\ -k_x \end{array}\right)
\end{align}
\begin{align}
\label{eq:OmegaDresselhaus}
\mathbf{\Omega}_{\k_1}^{\textrm{Dresselhaus}}=2\frac{\beta_D}{\hbar}\left(\begin{array}{c} k_y \\ k_x \end{array}\right)\, ,
\end{align}
\end{subequations}
for a (001)-grown quantum well with zinc-blende crystal structure. 
The spin-orbit interaction described by $\hat{H}_{SO}$ is responsible for
a precession of the electron spins in the effective magnetic field.

The momentum scattering is induced by the interaction between carriers and
localized Al impurities in the Al$_{x}$Ga$_{1-x}$As quantum well.
In contrast to the case of charged impurities as discussed, e.g., in 
Ref.~\onlinecite{NingDP}, Al ions are incorporated isoelectrically in the
GaAs matrix. 
Since the long-range part 
of the Coulomb interaction between Al impurities and the quasi-free carriers is
equally screened by the valence electrons as the long-range contribution of the
interaction between the carriers and the Ga ions that are replaced by Al ions,
the conduction band electrons experience only an effective short-range potential
at unit cells with Al ions, which can be described by the Hamiltonian \cite{nonmag}  
\begin{subequations}
\begin{align}
\label{eq:HImp}
\hat{H}_{\textrm{Imp}}=J\sum_{Ii}\delta\left(\mathbf{r}_i-\mathbf{R}_I\right).
\end{align}
The coupling constant is given by $J$, while $\mathbf{r}_i$ and $\mathbf{R}_I$ denote
the electron and impurity positions, respectively.  
In second quantization, $\hat{H}_{\textrm{Imp}}$ reads
\begin{align}
\label{eq:HImp2Q}
\hat{H}_{\textrm{Imp}}&=\frac{J}{V}\sum_{I=1}^N\sum_{\sigma\k\k'}e^{-i(\k'-\k)\cdot\R_I}\hat{c}_{\sigma\k}^\dagger\hat{c}_{\sigma\k'}^{}
\end{align}
\end{subequations}
with the system's volume $V$ and the number of impurity atoms $N$. 
In the following, we assume that the impurity positions are
determined by a fixed random distribution. This implies that the impurity
system is not changed by $\hat{H}_{\textrm{Imp}}$ and the scattering 
is elastic. Thus, $\hat{H}_{\textrm{Imp}}$ does not result in a thermalization 
as, e. g., momentum scattering due to carrier-phonon interactions.

We do not simulate the optical excitation of carriers via a
light-matter interaction Hamiltonian explicitly. Instead, we assume that an ultrashort
circularly polarized Gaussian pump pulse creates a spin polarized
electron distribution at $t\approx 0$. The corresponding optically excited
carrier distribution is then taken as an initial value for the
differential equations of motion. The validity of such a treatment has been previously verified for similar situations encountered in studies of diluted magnetic semiconductors\cite{SPIE}.

\subsection{Equations of Motion}
We are interested in the spin dynamics of the conduction band electrons 
in an Al$_{x}$Ga$_{1-x}$As quantum well, which can be obtained directly from the reduced electron density matrix
$\langle \hat{c}^\dagger_{\sigma_1\k_1} \hat{c}_{\sigma_2\k_1}\rangle$. 
Its time evolution is determined by the Heisenberg equation of motion
for the operator $\hat{c}^\dagger_{\sigma_1\k_1} \hat{c}_{\sigma_2\k_1}$.

While the effective single-particle Hamiltonians $\hat H_0$ and $\hat H_{SO}$
alone would yield a closed set of equations of motion for the reduced 
single-particle density matrix, the carrier-impurity interaction $\hat{H}_\textrm{Imp}$ is responsible for a
build-up of correlations between the electrons and impurities.
This can be seen most clearly by considering the time evolution of the
reduced density matrix due to the carrier-impurity interaction
\begin{align}
-i\hbar \ddt\big|_{\hat{H}_\textrm{Imp}} \langle  \hat{c}^\dagger_{\sigma_1\k_1} \hat{c}_{\sigma_2\k_1}\rangle &=\langle
[\hat{H}_\textrm{Imp}, \hat{c}^\dagger_{\sigma_1\k_1} \hat{c}_{\sigma_2\k_1}]\rangle.
\end{align}
Calculating the commutator and taking the average over the result 
yields terms of the form 
\begin{align}
\label{eq:origCor}
\langle e^{-i(\k_2-\k_1)\cdot \mathbf{R}_I} \hat{c}^\dagger_{\sigma_1\k_1} \hat{c}_{\sigma_2\k_2}\rangle,
\end{align}
which, in general, cannot be expressed in terms of the reduced density matrix 
alone since the averaging also involves taking an average over the random distribution of
the positions $\R_I$ of the impurities. 
Only for $\k_2=\k_1$, where 
$e^{-i(\k_2-\k_1)\cdot \mathbf{R}_I}\equiv 1$, the correlation in Eq. \eqref{eq:origCor}
reduces to $\langle \hat{c}^\dagger_{\sigma_1\k_1} \hat{c}_{\sigma_2\k_2}\rangle$.
In the spirit of Kubo's cumulant expansion \cite{Kubo62}, we subtract the
uncorrelated (mean-field) part of the term in Eq. (\ref{eq:origCor}) 
and define for $\k_2\neq \k_1$ the cumulants or true correlations
\begin{align}
\label{eq:KorrelationDef}
&\delta\langle e^{-i(\k_2-\k_1)\cdot\R_I}
\hat{c}_{\sigma_1\k_1}^\dagger\hat{c}_{\sigma_2\k_2}^{}\rangle\nonumber:=\\
&\langle e^{-i(\k_2-\k_1)
\cdot\R_I}\hat{c}_{\sigma_1\k_1}^\dagger \hat{c}_{\sigma_2\k_2}^{}\rangle-\langle e^{-i(\k_2-\k_1)
\cdot\R_I}\rangle\langle \hat{c}_{\sigma_1\k_1}^\dagger \hat{c}_{\sigma_2\k_2}^{}
\rangle.
\end{align}
Thus, the reduced electron density matrix is driven by carrier-impurity correlations. Similarly, the equations of motion for the carrier-impurity
correlations contain terms of the form 
\begin{align}
\label{eq:higherCor}
\langle e^{-i(\k-\k_2)\cdot\R_I}e^{-i(\k_2-\k_1)\cdot\R_{I'}}
\hat{c}_{\sigma_1\k_1}^\dagger\hat{c}_{\sigma_2\k}^{}\rangle.
\end{align}
For $I'=I$, $\k=\k_2$ or $\k_2=\k_1$, this expression reduces to the 
carrier-impurity correlations defined in Eq. (\ref{eq:origCor}) or the
carrier density matrix $\langle \hat{c}^\dagger_{\sigma_1\k_1} \hat{c}_{\sigma_2\k_1}\rangle$.
In the remaining cases, the cumulant expansion\cite{Kubo62} for three 
commuting random variables $A=e^{-i(\k-\k_2)\cdot\R_I}$, 
$B=e^{-i(\k_2-\k_1)\cdot\R_{I'}}$ and
$C=\hat{c}_{\sigma_1\k_1}^\dagger\hat{c}_{\sigma_2\k}^{}$ can be applied
\begin{align}
\label{eq:ABC}
\langle ABC\rangle=&
\delta\langle ABC\rangle 
+\langle A\rangle\delta\langle B C\rangle
+\langle B\rangle\delta\langle A C\rangle\nn&
+\langle C\rangle\delta\langle A B\rangle
+\langle A\rangle\langle B\rangle\langle C\rangle.
\end{align}
Here, we neglect higher order correlations involving different impurity positions
$\R_I$ and $\R_{I'}$, 
so that $\delta\langle ABC\rangle=\delta\langle A B\rangle=0$. The remaining terms
involve either $\langle A\rangle$ or $\langle B\rangle$. Assuming an on average
homogeneous impurity distribution \cite{Thurn:12} one obtains
\begin{align}
\label{eq:Isotropie}
\langle e^{-i\k\cdot\R_I}\rangle&=\delta_{\k0}\, 
\end{align}
and therefore $\langle A\rangle=\delta_{\k,\k_2}$ as well as $\langle B\rangle=
\delta_{\k_2,\k_1}$. Thus, all terms in  Eq. \eqref{eq:ABC} vanish, except for
those with $\k=\k_2$ or $\k_2=\k_1$, which can be expressed by the
lowest-order cumulants defined in Eq. \eqref{eq:KorrelationDef} and 
the carrier density matrix. 

This way, a closed set of equations of motion is obtained for the
dynamical variables
\begin{subequations}
\label{eq:Def_CCq}
\begin{align}
\label{eq:Def_C}
C_{\sigma_1\k_1}^{\sigma_2}&:=\langle\hat{c}_{\sigma_1\k_1}^\dagger
\hat{c}_{\sigma_2\k_1}^{}\rangle\\
\label{eq:Def_Cq}
\overline{C}_{\sigma_1\k_1}^{\sigma_2\k_2}&:=
V\delta\langle e^{-i(\k_2-\k_1)\cdot\R_I}\hat{c}_{\sigma_1\k_1}^\dagger
\hat{c}_{\sigma_2\k_2}^{}\rangle,
\end{align}
\end{subequations}
where $C_{\sigma_1\k_1}^{\sigma_2}$ is the electron density matrix and
$\overline{C}_{\sigma_1\k_1}^{\sigma_2\k_2}$ are the carrier-impurity
correlations, where the latter are only defined for $\k_2\neq\k_1$. 
It is convenient to rescale the correlations by the factor $V$ so that they remain finite
in the limit $V\to\infty$.
Note that a similar correlation expansion has been developed and applied in Refs. 
\onlinecite{AxtExcForm,RandPotZimmermann,RandPotZimmermann2,RandPotZimmermann3},
e.g., for investigations of the influence of interface roughness on exciton 
line shapes.

The quantum kinetic equations of motion for the dynamical variables are
\begin{subequations}
\label{eq:DGL_CCq}
\begin{align}
\label{eq:DGL_C}
-i\hbar\ddt C_{\sigma_1\k_1}^{\sigma_2}
&=\hbar\mathbf{\Omega}_{\k_1}\cdot\sum_{\sigma}\left(\mathbf{s}_{\sigma\sigma_1}C_{\sigma\k_1}^{\sigma_2}
-\mathbf{s}_{\sigma_2\sigma}C_{\sigma_1\k_1}^{\sigma}\right)\nonumber\\
&+\frac{JN}{V^2}\sum_{\k\neq\k_1}\left(\overline{C}_{\sigma_1\k}^{\sigma_2\k_1}-\overline{C}_{\sigma_1\k_1}^{\sigma_2\k}\right)\\
\label{eq:DGL_Cq}
-i\hbar\ddt\overline{C}_{\sigma_1\k_1}^{\sigma_2\k_2}
&=\hbar\left(\omega_{\k_1}-\omega_{\k_2}\right)\overline{C}_{\sigma_1\k_1}^{\sigma_2\k_2}\nonumber\\
&+\sum_{\sigma}\hbar\left(\mathbf{\Omega}_{\k_1}\cdot\mathbf{s}_{\sigma\sigma_1}\overline{C}_{\sigma\k_1}^{\sigma_2\k_2}
-\mathbf{\Omega}_{\k_2}\cdot\mathbf{s}_{\sigma_2\sigma}\overline{C}_{\sigma_1\k_1}^{\sigma\k_2}\right)\nonumber\\
&+\frac{J}{V}\left(\sum_{\k\neq\k_1}\overline{C}_{\sigma_1\k}^{\sigma_2\k_2}
-\sum_{\k\neq\k_2}\overline{C}_{\sigma_1\k_1}^{\sigma_2\k}\right)\nonumber\\
&+J\left(C_{\sigma_1\k_2}^{\sigma_2}-C_{\sigma_1\k_1}^{\sigma_2}\right).
\end{align}
\end{subequations}

The first term on the r.h.s. of Eq. (\ref{eq:DGL_C}) describes 
the mean field precession of the electron spins in the effective field
while the second term incorporates the changes of the 
electron density matrix due to the carrier-impurity correlations that
mediate the impurity scattering.
The equation of motion (\ref{eq:DGL_Cq}) for the correlations has the 
structure of an oscillator with a frequency corresponding to
the difference in kinetic energies $\hbar\omega_{\k_2}-\hbar\omega_{\k_1}$
(first term on the r.h.s.) driven by the electron density matrix 
$C_{\sigma_1\k_1}^{\sigma_2}$ via the last term on the r.h.s. of 
Eq. (\ref{eq:DGL_Cq}). The second term describes the precession of 
the carrier-impurity correlations around the effective magnetic field and
the third term accounts for changes of the wave vectors of the 
correlations caused by the carrier-impurity interaction.

\subsection{Markov Limit}
The full quantum kinetic equations of motion (\ref{eq:DGL_CCq}) describe 
a dynamics of the electron density matrix that is non-Markovian in general,
i.e., the correlations induce a finite memory. 
It is instructive to consider the Markovian limit of the quantum kinetic equations because of two reasons:
On the one hand, to investigate the importance of finite-memory effects, and on
the other hand, to derive an analytic expression for the momentum scattering 
rate, so that our theory can be related to more commonly used approximate descriptions 
of the DP mechanism.

The Markov limit of the quantum kinetic equations of motion is obtained by neglecting
the second and third terms on the r.h.s. of Eq. (\ref{eq:DGL_Cq}), which allows one to
formally integrate the correlations yielding 

\begin{align}
\label{Markov_Herleitung_1}
&\overline{C}_{\sigma_1\k_1}^{\sigma_2\k_2}(t)=e^{i\left(\omega_{\k_1}-\omega_{\k_2}\right)t}\Big(\overline{C}_{\sigma_1\k_1}^{\sigma_2\k_2}(t_0)\nonumber\\
&+\int_{t_0}^tdt'i\frac{J}{\hbar}\left(C_{\sigma_1\k_2}^{\sigma_2}(t')-C_{\sigma_1\k_1}^{\sigma_2}(t')\right)e^{-i\left(\omega_{\k_1}-\omega_{\k_2}\right)t'}\Big)\, .
\end{align}
Neglecting the respective terms in the equation for the correlations can be justified 
as follows: the third term in Eq. (\ref{eq:DGL_Cq}) is a higher order term with 
respect to the coupling constant. Furthermore, it consists of a sum of correlations 
with different wave vectors that oscillate with different frequencies and can therefore
be expected to dephase fast.
The second term in Eq. (\ref{eq:DGL_Cq}) mainly accounts for the fact that 
the energy eigenvalues and eigenstates of the semiconductor crystal, 
which define the electronic states between which elastic momentum
scattering events take place, are modified by the effective magnetic field.
Here, however, we mainly consider situations where the spin-orbit
splitting of the conduction subbands $\hbar\Omega_\k$ 
is on average smaller than the average kinetic energy and the modification
of the band structure due to the effective field is of minor importance.
Situations where this modification is important have been discussed in
Ref.~\onlinecite{Grimaldi2005} on the level of a Markovian theory. 
Note that it is also possible to formally integrate Eq. (\ref{eq:DGL_Cq}) 
accounting for the second term on the r.h.s., but the resulting Markovian equations
become more involved \footnote{A similar problem has been described and
solved in Ref.~\onlinecite{kdep}, where the spin dynamics in diluted magnetic
semiconductors in the presence of a $\k$-dependent magnetic field was studied}.
An \emph{a posteriori} justification for neglecting the respective terms 
in the equations of motion will be given by comparing numerical calculations 
of the full quantum kinetic equations and the Markovian equations.

The Markovian approximation is characterized by the assumption of a short memory,
which implies that the density matrices $C_{\sigma_1\k_1}^{\sigma_2}(t')$ in
Eq. (\ref{Markov_Herleitung_1}) can be evaluated at $t'=t$ and the lower limit of
the memory integral can be extended to $t_0\to-\infty$ \cite{RossiKuhn02,PESC}.
Finally, using $\int_{-\infty}^0 dt\,e^{-i\Delta\omega t}=
\pi\delta(\Delta\omega)+\mathcal{P}\frac{i}{\Delta\omega}$
and assuming that the correlations are initially zero $\overline{C}_{\sigma_1\k_1}^{\sigma_2\k_2}(t_0\to-\infty)=0$,
we obtain
\begin{align}
\label{eq:DGL_Markov}
-i\hbar&\ddt C_{\sigma_1\k_1}^{\sigma_2}
=\hbar\mathbf{\Omega}_{\k_1}\cdot\sum_{\sigma}\left(\mathbf{s}_{\sigma\sigma_1}C_{\sigma\k_1}^{\sigma_2}
-\mathbf{s}_{\sigma_2\sigma}C_{\sigma_1\k_1}^{\sigma}\right)\nonumber\\
&+2\pi i\frac{J^2N}{V^2}\sum_{\k\neq\k_1}
\left(C_{\sigma_1\k_1}^{\sigma_2}-C_{\sigma_1\k}^{\sigma_2}\right)
\delta\left(\hbar\omega_{\k}-\hbar\omega_{\k_1}\right)\, .
\end{align}
The contributions from the principal value cancel exactly.
Equation~\eqref{eq:DGL_Markov} can be rewritten in the quasi-continuous limit $\sum\limits_\k \to 
\int d(\hbar\omega_\k) D^{2D}(\hbar\omega_\k)$ with 
the two-dimensional spectral density of states 
$D^{2D}(\hbar\omega)= \frac{Am^*}{2\pi \hbar^2}$
in terms of the more intuitive variable 
$\langle\mathbf{s}_{\k_1}\rangle=\sum\limits_{\sigma\sigma'}\mathbf{s}_{\sigma\sigma'}C_{\sigma\k_1}^{\sigma'}$,
i.e., the average spin in the electronic states with wave vector $\k_1$.
We obtain 
\begin{align}
\label{eq:DGL_Markov_sk}
\ddt\langle\mathbf{s}_{\k_1}\rangle
&=\mathbf{\Omega}_{\k_1}\times\langle\mathbf{s}_{\k_1}\rangle
-\frac 1{\tau_p}\big(\langle \mathbf{s}_{\k_1}\rangle
-\langle \overline{\mathbf{s}}_{k_1}\rangle \big),
\end{align}
with the momentum scattering rate 
\begin{align}
\label{eq:taup}
\frac 1{\tau_p}=\frac{4J^2 m^* x}{\hbar^3da^3}\, ,
\end{align}
where $x$ denotes the impurity concentration, $d$ the thickness of the quantum well, and $a$ the lattice constant of the crystal.
The average spin in the shell of states with modulus $k_1$ of the wave vector $\k_1$ is given by
\begin{align}
\langle \overline{\mathbf{s}}_{k_1}\rangle=&
\frac 1{2\pi}\int \limits_0^{2\pi} d\varphi\; 
\langle \mathbf{s}_{\k(k_1,\varphi)}\rangle,
\end{align}
where
\begin{align}
\k(k_1, \varphi)=&\left(\begin{array}{c} 
k_1\cos(\varphi)\\ k_1\sin(\varphi)
\end{array}\right).
\end{align}
Thus, in the Markov limit, the time evolution of the electron spins is given
by a precession of the spin in the $\k$-dependent magnetic field and a
redistribution of electron spins within a shell of 
fixed kinetic energy $\hbar\omega_{\k_1}$ with the momentum scattering rate
$\tau_p^{-1}$.
Note that the Markovian equation of motion \eqref{eq:DGL_Markov_sk} can also be
derived by other approaches \cite{Gridnev2001} and they are, in fact, 
also applicable in settings where a phenomenological momentum scattering rate is used
to incorporates additional effects due to, e.g., carrier-carrier or carrier-phonon scattering.
Thus, the results obtained from the Markovian equations are valid in more 
general DP scenarios where the momentum scattering does not need to be 
caused by localized impurities. 

\subsection{Limiting cases}
\label{subsec:Limiting-cases}
Eqs. (\ref{eq:DGL_CCq}) and (\ref{eq:DGL_Markov_sk}) 
describe the time evolution of an ensemble of optically induced electron spins. 
In contrast, the conventional description of the DP mechanism, which can be used to
derive analytic expressions for the spin relaxation time in limiting cases,
is based on a different picture where a single electron is 
considered which performs a stochastic motion in $\k$-space. \cite{DP,WuReview,DPEYunified}
Here, we review the basic results of this stochastic picture.

In the conventional DP picture, it is assumed that the electron's 
wave vector $\k$ changes randomly after a 
time interval corresponding to a correlation time $\tau_c$, which we identify
with the momentum scattering time $\tau_p$ defined in Eq. (\ref{eq:taup}). 
During this correlation time, the electron spin precesses about 
the effective field $\boldsymbol\Omega_\k$. 
Thus, between each scattering event the electron spin
changes about an angle of $\theta_\k =\tau_c\Omega_\k$.
If the angle $\theta_\k$ is small, which implies that the scattering 
rate $\tau_c^{-1}$ is much larger than the typical precession frequency,
the time evolution of the electron spin can be regarded 
as a random walk consisting of $n=t/\tau_c$ time steps. 
The root mean square of the precession angle is then given 
by 
\begin{align}
\sqrt{\Delta \theta^2}&=\sqrt{ \langle \theta_\k^2 \rangle \frac t{\tau_c}}=
\sqrt{\langle \Omega_\k^2\rangle \tau_c t},
\end{align}
which is of the order of unity at the spin relaxation time 
\begin{align}
\label{eq:tau_s_DP}
\tau_s\sim\frac 1{\langle \Omega_\k^2\rangle \tau_c},
\end{align}
where the brackets
indicate the average over the $\k$-space states available for the
random walk process.
This way, one obtains the well-known DP result that the spin relaxation
time is predicted to be inversely proportional to the momentum
relaxation time. 

As stated above, the derivation of the expression for the spin relaxation 
time $\tau_s$ in Eq. (\ref{eq:tau_s_DP}) requires the assumption of the 
strong scattering limit $\tau_c^{-1}\gg \sqrt{\langle\Omega_\k^2\rangle}$.
However, analytic expressions for the spin relaxation time can also be
obtained in the opposite limit, $\tau_c^{-1}\ll \sqrt{\langle\Omega_\k^2\rangle}$.
\cite{Gridnev2001} 
This can be done by starting from Eq. (\ref{eq:DGL_Markov_sk})
and considering an initial carrier spin polarization along the 
$z$-axis (growth direction). First, the $z$-component of
Eq. (\ref{eq:DGL_Markov_sk}) is differentiated. In the resulting equation, the 
expressions $\langle s^x_{\k_1}\rangle$ and $\langle s^y_{\k_1}\rangle$ 
have to be eliminated by expressing them in terms of $\langle s^z_{\k_1}\rangle$ 
and $\ddt \langle s^z_{\k_1}\rangle$ using again Eq. (\ref{eq:DGL_Markov_sk}).
If terms higher than first order in the momentum scattering rate $\tau_p^{-1}$ 
are neglected and if it is assumed that the modulus of the precession frequency is
independent of the polar angle of $\k_1$, i.e., one can write
$\sqrt{\boldsymbol\Omega^2_{\k_1}}=\Omega_{k_1}$, 
one obtains the second order differential equation 
for the $z$-component of the average electron spin
\begin{align}
\label{eq:low_scatt_dgl}
\frac{\partial^2}{\partial t^2} \langle \overline{s}^z_{k_1}\rangle
+\frac{1}{\tau_p}  \ddt \langle \overline{s}^z_{k_1}\rangle
+\Omega_{k_1}^2 \langle \overline{s}^z_{k_1}\rangle = 0.
\end{align}
Eq. (\ref{eq:low_scatt_dgl}) has the form of a damped oscillator whose solution 
for $\frac{1}{\tau_p}\ll\Omega_{k_1}$ is an oscillation with frequency
$\Omega_{k_1}$ that decays exponentially with the relaxation rate
\begin{align}
\label{eq:tau_s_ws}
&\frac 1{\tau_s}=\frac 1{2\tau_p}.
\end{align}
Thus, Eq. (\ref{eq:tau_s_ws}) predicts that, in the weak-scattering limit, 
the spin relaxation rate is proportional to the momentum relaxation rate.

It is noteworthy that, in contrast to the strong-scattering limit, 
the weak-scattering limit supports oscillations of the spin polarization. 
Thus, when an ensemble of electrons with different precession frequencies
is considered, the superposition of the different oscillations may 
additionally lead to a dephasing, which causes a decay of the total
electron spin even in the absence of momentum scattering $\tau_p^{-1}=0$.
This effect has been described by Ning \emph{et al.}  
as an inhomogeneous broadening mechanism and was explored numerically in
Refs.~\onlinecite{WuNingEuroPhys,NingDP}.

In situations where the electron occupation is well described
by a quasi-equilibrium Fermi distribution with a significant Fermi energy,
such as in n-doped systems or in transport experiments, only
electronic states with a wave vector close to the Fermi wave vector 
$|\k|\approx k_F$ are relevant for the spin dynamics. Because the 
modulus $|\k|$ of the wave vector $\k$ determines the precession frequency
$|\boldsymbol\Omega_\k|$, there is essentially only one precession frequency
present in these situations and no DIIB takes place.

In contrast, in the case of an optically induced spin-polarized electron
distribution in an intrinsic semiconductor, the spectral width of the
exciting laser, e.g., due to the energy-time uncertainty associated with the finite 
duration of the laser pulse, gives rise to a corresponding finite spectral width 
of the electron distribution. In general, this translates into a
non-negligible width of the distribution of the modulus $|\k|$ of the wave
vectors of spin polarized carriers.
Therefore, here, the DIIB can be expected to 
be relevant. \cite{Proceedings_Pablo_2015}

The qualitative shape of the time evolution of the total electron spin
due to the DIIB alone, i.e., in the absence of 
momentum scattering $\tau_p\to \infty$, can be
discussed in some limiting cases.
First, consider the case 
of the initial distribution of spin polarized electrons defined by 
\begin{subequations}
\begin{align}
&n_\k(t=0)=
\begin{cases} n_0\,, &
|\k|\in [ k_0 -\frac 12 \Delta k ; k_0 +\frac 12 \Delta k]\\
0\,, & \textrm{else}
\end{cases}\\
&s^z_{\k}(t=0)=\frac 12 n_\k(t=0),
\end{align}
i.e., a distribution centered at $k_0$ with width $\Delta k$.
Additionally, assume for simplicity that the system is only subject 
to Rashba spin-orbit interaction, so that 
\begin{align}
s^z_\k(t)\approx \frac {n_0}2\cos\big(2\frac{\alpha_R}\hbar k t\big).
\end{align}
If $\Delta k\ll k_0$, we can assume that the 
two-dimensional $k$-dependent density of states is approximately constant
$D^{2D}(k)\approx D^{2D}(k_0)$ and that the spin decay rate is essentially 
independent of $\k$. Then, the total spin is given by
\begin{align}
&\int dk \, D^{2D}(k) s^z_\k(t)\approx\frac{n_0}2 D^{2D}(k_0)
\int\limits_{k_0 -\frac 12 \Delta k}^{k_0 +\frac 12 \Delta k} dk\,
\cos\Big(2\frac{\alpha_R}\hbar k t\Big)\nn
&= \frac{n_0}2 D^{2D}(k_0) 
\frac{ \sin\big[\frac{2\alpha_R (k_0 +\frac 12 \Delta k)t}\hbar\big]
-\sin\big[\frac{2\alpha_R (k_0 -\frac 12 \Delta k)t}\hbar\big]}
{ 2\frac{\alpha_R}\hbar t}.
\end{align}
\end{subequations}
Thus, in this situation, we find that the total electron spin decays
algebraically as $\frac 1t$. Note that, 
because this behavior deviates strongly from an exponential decay, it is not
possible to unambiguously associate a spin decay time with the spin dynamics.

In Ref.~\onlinecite{Proceedings_Pablo_2015}, another limiting case of the
DIIB was discussed where a Gaussian initial spin-polarized
spectral electron distribution was considered. It was found that, when the
Gaussian distribution is centered around the band edge, i.e., $k\approx 0$, 
the time evolution of the total electron spin is given by an expression that
resembles a one-sided Fourier-transform of a function with a Gaussian-like shape.
Thus, the time evolution of the total electron spin itself is expected to
be well approximated by a Gaussian rather than an exponential. 
This expectation was supported by numerical calculations in 
Ref.~\onlinecite{Proceedings_Pablo_2015}.

To summarize, in the Markov limit,
an explicit expression for the spin relaxation rate $\tau_s^{-1}$ 
can be given in the strong-scattering limit 
$\tau_p^{-1}\gg \sqrt{\langle\Omega_\k^2\rangle}$ by Eq.~(\ref{eq:tau_s_DP}), 
which is the original result of D'yakonov and Perel' and which predicts
a spin relaxation rate inversely proportional to the momentum scattering rate.
In the weak-scattering limit $\tau_p^{-1}\ll \sqrt{\langle\Omega_\k^2\rangle}$,
the total spin in a single shell of states with fixed wave vector modulus
$|\k|$ decays exponentially, where the spin relaxation rate is one half of the
momentum relaxation rate. If, however, a distribution of spin-polarized 
electrons with varying values of $|\k|$ is optically 
excited, the DIIB mechanism predicts an algebraic or a
Gaussian decay of the initial electron spin, depending on the spectral 
properties of the initial electron distribution.

\section{Results}
Before presenting the results of numerical calculations, we first discuss the 
parameters used in our study as well as the details of the numerical methods
used for the calculations.
\begin{figure*}
	\includegraphics[width=16cm,height=6cm]{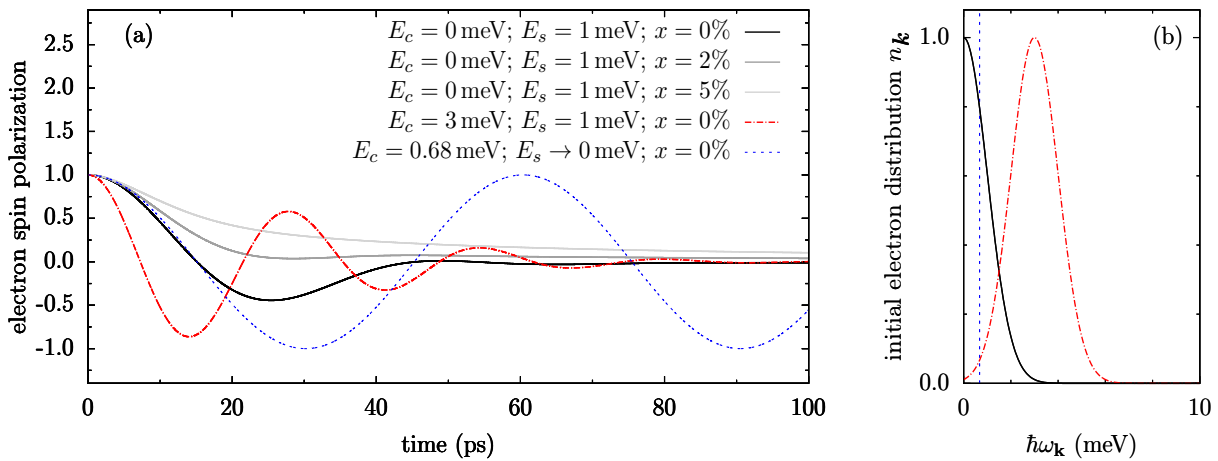}
	\caption{\label{fig:Dyn} (a): Time evolution of the optically induced 
electron spin calculated using the Markovian equation (\ref{eq:DGL_Markov_sk})
for different impurity concentrations $x$ and different initial electron
distributions (Gaussian with central energy $E_c$ above the band edge and
standard deviation $E_s$) shown in (b).
}
\end{figure*}

\subsection{System Parameters}
In this article, we study the spin dynamics in a narrow 
Al$_{x}$Ga$_{1-x}$As quantum well immediately
after optical excitation with circularly polarized light.
The Al content $x$ in the quantum well determines the 
momentum scattering and will be varied from zero to a 
few percent.
Furthermore, we assume that the crystal can be well described by a zinc-blende
lattice with parameters close to that of GaAs. For our calculations, we use
the lattice constant $a=565.35\,\textrm{pm}$ and the effective conduction band 
electron mass $m^*=0.0665\cdot m_0$, where $m_0$ is the free electron mass.

The coupling constant $J$ is chosen in such a way \cite{nonmag}
that it is, on a mean-field level,
consistent with the conduction band offset at a GaAs/Al$_{x}$Ga$_{1-x}$As 
interface of $\Delta E_c=x\cdot 0.87\,\textrm{eV}$ in magnitude\cite{Yi2009}. 
From this consideration, we obtain the coupling constant 
$J=\frac{a^3}{4}\Delta E_c=39\,\textrm{meV}\textrm{nm}^3$.

We choose the width of the quantum well to be $d=10\,\textrm{nm}$ and only
consider the lowest confinement state, for which 
$\langle k_z^2\rangle=\left(\pi/d\right)^2$. Then, the Dresselhaus parameter 
is given by $\beta_D=-\gamma\langle k_z^2\rangle$ with 
$\gamma=-11\,\textrm{meV}\textrm{nm}^3$ (cf. Ref~\onlinecite{Ganichev14})
yielding $\beta_D\approx 1\,\textrm{meV}\textrm{nm}$.

The Rashba coefficient on the other hand is not only dependent on the material, 
but also on external electric fields and potentials \cite{Ganichev14}. 
We regard the Rashba coefficient as a tunable parameter and, for the sake of 
simplicity, set it to $\alpha_R=0$, if not mentioned otherwise.

The optical driving is modeled by choosing suitable initial values for the
electron density matrix. We imagine that a single circularly polarized Gaussian 
ultrafast femtosecond laser pulse has selectively excited spin-up electrons at $t=0$ via the
spin selection rules. Consistent with the spectral properties of such a pump pulse,
we assume that at $t=0$ only the spin-up occupations in the electron density matrix
are populated and the spectral electron density is described by a Gaussian centered 
at an energy $E_c$ above the band edge with a spectral standard deviation $E_s$.

\subsection{Numerical methods}

With the initial values described above, we numerically solve either the full 
quantum kinetic equations (\ref{eq:DGL_CCq}) or the Markovian equation 
(\ref{eq:DGL_Markov_sk}) using a fourth-order Runge-Kutta algorithm. 
In order to arrive at a numerically tractable
problem, we only consider electronic states up to 
a cut-off energy of about $20$ meV. 
Furthermore, we 
take the quasi-continuous limit and replace sums over $\k$ with the
corresponding two-dimensional $\k$-space integral, which is then treated in 
polar coordinates. The quantities depending on the polar angle of a wave vector 
are then expanded in terms of a discrete Fourier-series, 
which turns out to drastically speed up the calculations.
This procedure makes it possible to equally treat all directions in 
$\k$-space, whereas in other approaches \cite{NingDP} only selected directions,
e.g., the coordinate axes, could be resolved.
The modulus $|\k|$ of the wave vector is discretized straightforwardly.
It has been checked that neither refining the discretization of the $\k$-space and the
time discretization nor increasing the cut-off energy further 
leads to visibly different results from those presented below.

\subsection{Time evolution}

We now discuss the general features of the time evolution of the electron spin
polarization as shown in Fig. \ref{fig:Dyn}(a), where the spin polarization 
is defined by $\sum_\k \langle s^z_\k\rangle/(\frac 12N_e)$ with 
total electron number $N_e=\sum_{\k \sigma} C_{\sigma\k}^{\sigma}$.

First of all, for low impurity concentrations and therefore
low momentum scattering rates, pronounced oscillations of the total electron 
spin are found, whereas with increasing impurity concentration, the oscillations
are suppressed and the electron spin polarization eventually decays monotonically.
It can be seen from the results presented in Fig. \ref{fig:Dyn}(a) that 
the time evolution of the spin polarization is, in general, not well described
by an exponentially damped oscillation, in particular for strong momentum
scattering. For example, the graph for $x=2\%$ shows non-monotonic behavior whilst displaying always positive spin polarization. In contrast, negative spin polarizations would be expected from an exponentially damped cosine.
It is noteworthy that the spin polarization decays even for $x=0$, 
where, according to the stochastic picture without DIIB,
the spin decay rate is expected to vanish, 
because $\tau_s^{-1}\approx \frac 12\tau_p^{-1}\to 0$.

In Fig. \ref{fig:Dyn}(a), also the spin dynamics for different 
initial electron distributions, which are shown in Fig. \ref{fig:Dyn}(b), 
is presented, corresponding to different properties of the exciting laser 
pulse. The center of the Gaussian $E_c$ measured from the band edge can be
controlled by the central frequency of the exciting laser and the width 
(standard deviation $E_s$) is related to the spectral properties of the laser 
pulse and has a lower bound due to the energy-time uncertainty.
Nevertheless, it is instructive to discuss the theoretical case of a
spectrally sharp initial spin-polarized carrier distribution with $E_s\to 0$, 
since this situation corresponds to turning off the effect of 
DIIB.
The calculated time evolution in such a situation is depicted by the
dotted line in Fig. \ref{fig:Dyn}(a). The center of the spectrally sharp
initial electron distribution is chosen in such a way that it has the
same average wave vector modulus $\langle |\k|\rangle$ as the initial
electron distribution of the Gaussian with $E_c=0$ meV and $E_s=1$ meV.
Comparing the corresponding calculations with zero and finite width of the
electron distributions reveals that the DIIB
is responsible for the spin decay in the absence of momentum scattering when 
the initial electron distribution has a finite width,
whereas the oscillations continue indefinitely 
in calculations with spectrally sharp initial electron distribution if $x=0$. 

When the center of the electron distribution is shifted to higher energies
(dashed-dotted line in Fig. \ref{fig:Dyn}), the oscillation frequency is also 
increased. This can be explained by the fact that the strength of the
Rashba field $\Omega_\k$, which determines the typical precession frequency
in the system, increases with increasing wave vector modulus $|\k|$ or, 
equivalently, increasing kinetic energy $\hbar\omega_\k$.
Furthermore, the shift of the center of the electron distributions to higher
energies leads to a reduction of the spin decay. We attribute this to the fact
that for a Gaussian distribution with $E_c \gg E_s$, the situation 
resembles that of a spectrally sharp distribution and DIIB
becomes less important.

\subsection{Dependence of the spin relaxation times on momentum scattering}
\label{subsec:0meV}

\begin{figure*}
		\includegraphics[width=\textwidth]{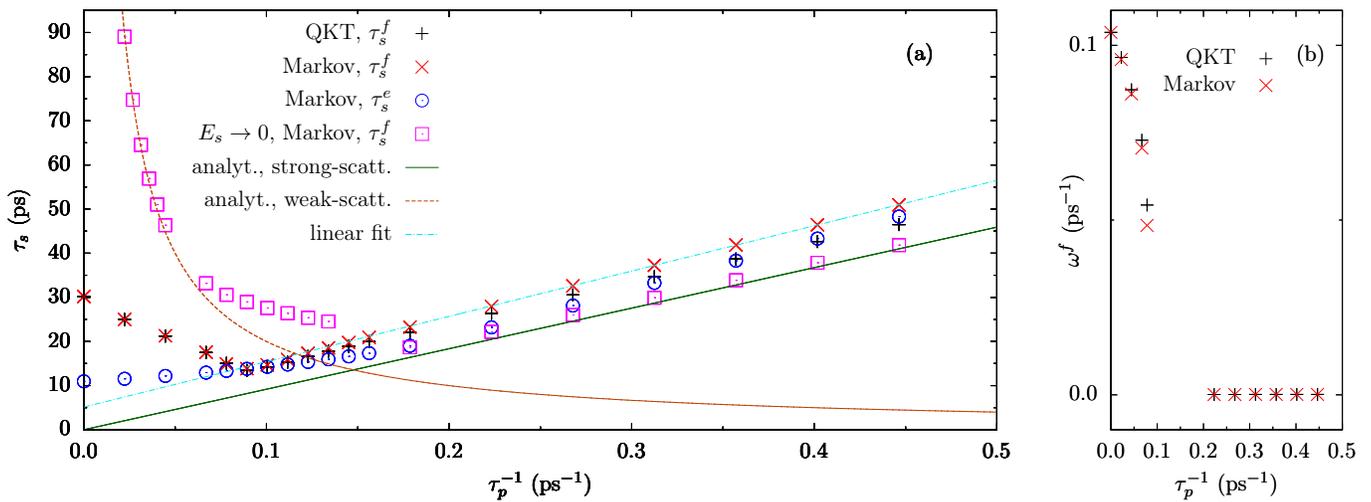}
		\caption{\label{fig:rates0} (a) Spin decay time $\tau_s$ 
		as a function of the momentum scattering rate $\tau_p^{-1}$.
		The spin decay times are determined either by fitting a stretched
		exponential to the time evolution of the total spin polarization 
		($\tau_s^f$) or by extracting the time after which the spin polarization
		is decayed to $\frac{1}{e}$ of its initial value ($\tau_s^e$).
		The black pluses represent the results of calculations using the full
		quantum kinetic equations, while the remaining results are based on the
		Markovian equations of motion. The initial electron distributions used 
		for the calculations are Gaussians with standard deviation $E_s=1$ meV 
		($E_s\to 0$ for the results depicted as purple squares) centered at 
		$E_{\textrm{c}}=0\,\textrm{meV}$
		[cf. Fig. \ref{fig:Dyn}(b)].
		The analytic expressions in the strong- and weak-scattering
		limits are depicted as lines. For comparison, a linear fit through the
		last five points of the Markovian calculation of 
		$\tau_s^f$ is shown.
		(b) Precession frequency of the total spin polarization obtained 
		from the fitting procedure. 
		}
\end{figure*}

It is common in the literature \cite{DP, Gridnev2001,WuReview} to discuss 
spin relaxation times or rates and their dependencies on different 
parameters. However, as we have seen in Fig. \ref{fig:Dyn}(a), the
spin dynamics can strongly deviate from an exponential behavior that is
implied by the concept of a spin relaxation rate. Thus, 
the spin relaxation rate becomes ill-defined and ambiguous in certain 
cases. 

Nevertheless, it is useful for understanding the qualitative dependence of
the spin dynamics on the model parameters to consider quantities that can, 
to a certain extent, be interpreted as a characteristic time for the decay
of the spin polarization. Here, we discuss two different 
definitions of spin decay times. 

First, we fit a stretched exponential of the form 
\begin{align}
f(t)=\cos(\omega^f t)\exp[-(t/\tau^f_s)^n]
\end{align}
to the time evolution of
the spin polarization, where $\omega^f$, $\tau^f_s$, and $n$ are free parameters. 
The value of $\tau_s^f$ is then considered to be a measure of the spin decay time. 
The variable parameter $n$ in the stretched exponential 
allows one to extract a meaningful spin decay time, e.g., in the limiting cases 
where an exponential or a Gaussian decay is expected.
Second, we define $\tau_s^e$ to be the time after which the spin polarization has 
decreased to a value of $\frac{1}{e}$ of its initial value.

The parameters $\tau_s^e$ and $\tau_s^f$ are, in general, not equivalent.
For example, $\tau_s^e$ is, in general, smaller than $\tau_s^f$
since also the oscillatory part $\cos(\omega^f t)$ leads to a decay of the 
total signal for small times.
The different aspects of the spin dynamics measured by $\tau_s^e$ and $\tau_s^f$
can be discussed, e.g., for the time evolution of the spin polarization for
$x=2\%$ in Fig. \ref{fig:Dyn}(a). There, the spin polarization first decays rapidly,
then it increases again slightly and eventually decays very slowly toward zero.
In this situation, the initial fast decay is measured by $\tau_s^e$, while the
slow decay at long times enters via the fit procedure in $\tau_s^f$, which therefore
measures the overall time scale of the spin decay.  

In Fig. \ref{fig:rates0}(a), the spin relaxation times
 $\tau_s^e$ and $\tau_s^f$ obtained from calculations of the spin dynamics
 are depicted as a function of the momentum scattering rate $\tau_p^{-1}$ determined from eq. (\ref{eq:taup}), which
is varied by changing the impurity concentration $x$ in the calculations.
For comparison, the analytic results in the strong-scattering 
(solid straight line) and the weak-scattering (dashed hyperbola) limits 
are also depicted. It is found that above $\tau_p^{-1}\approx 0.1$ ps$^{-1}$ 
both definitions of the spin decay times $\tau_s^f$ and $\tau_s^e$
lead to quantitatively different results, but depend qualitatively on 
the momentum scattering in a similar way and 
follow the general trend expected in the strong scattering limit. 
However, the numerically obtained
spin decay times are consistently larger than the DP result, even for the largest studied
momentum scattering rate. This tendency is visualized by fitting a line 
through the last five points of the Markovian result for 
$\tau_s^f$. 

Let us first concentrate on the Markovian results. For momentum scattering rates below $0.1$ ps$^{-1}$, the 
results of $\tau_s^e$ and $\tau_s^f$ differ significantly:
The spin decay time $\tau_s^e$, which measures the fast initial decay,
decreases monotonically with decreasing momentum scattering rate. 
But $\tau_s^f$, which measures the overall decay of the spin polarization including the 
long-time parts, shows a pronounced kink and a minimum at 
$\tau_p^{-1}\approx 0.1$ ps$^{-1}$.
The discrepancy between $\tau_s^e$ and $\tau_s^f$ can be traced back 
to the fact that, for small momentum scattering rates,
the time evolution of the spin oscillates and, as explained above, the
oscillatory part leads to a decay of the spin that is included in the decay
time $\tau_s^e$ but not in $\tau_s^f$.
The momentum-scattering-dependence
of the precession frequency $\omega^f$ obtained by the fitting procedure 
is presented in Fig. \ref{fig:rates0}(b) and supports this explanation. 
The results depicted in Fig. \ref{fig:rates0}(b) indicate a bifurcation 
point close to the kink in $\tau_s^f$ in Fig. \ref{fig:rates0}(a), below 
which oscillations occur. 
However, for values close to $\tau_p^{-1}=0.1$ ps$^{-1}$, i.e., the region of the
kink and the onset of the oscillations, the fitting procedure does not 
produce reliable results for $\omega^f$, as small changes in the initial values 
of the fitting parameters can lead to significantly different results.
Thus, in Fig. \ref{fig:rates0}(b), we present only values for $\omega^f$ which 
are stable with respect to changes in the initial 
values of the fit parameters, which excludes the region of the expected 
bifurcation point.
 
It is noteworthy that the numerical results for the spin decay times
for $\tau_p^{-1}\to 0$ disagree quantitatively and qualitatively with the
analytical result $\tau_s=2\tau_p$. In particular, the numerically obtained
spin decay time $\tau_s^f$ increases approximately linearly to a finite value when 
$\tau_p^{-1}\to 0$, whereas the analytical result predicts a divergence, i.e., the
spin decay time becomes infinitely long. However, as discussed in section \ref{subsec:Limiting-cases}, 
DIIB can become important in the limit $\tau_p^{-1}\to 0$.
To investigate the influence of DIIB, we 
present in Fig. \ref{fig:rates0}(a) (purple squares)
also the spin decay time $\tau_s^f$ 
obtained from calculations with a spectrally sharp initial electron distribution
with the same value of $\langle |\k|\rangle$ as the Gaussian distribution
used for the calculations discussed so far. 
It can be seen that the results of these simulations 
coincide with the analytical results in the strong- and weak-scattering limits.

Thus, the discrepancies between numerical calculations for the Gaussian 
electron distribution and the analytical results in the respective limits can be
traced back to the finite width of the spectral electron distribution.
In the weak-scattering limit, the DIIB becomes
important and dominates the spin decay. In the strong-scattering limit,
the finite spectral width is found to increase the spin decay time, i.e., 
the spin decay is reduced. The reason for this is that the spin decay time 
according to the DP result given by Eq. \eqref{eq:tau_s_DP} is inversely proportional to
$\langle\Omega^2_{\k_1}\rangle$ and the spin relaxes faster in states with
larger wave vectors. Because of the finite width of the electron distribution,
the ensemble of electron spins has parts whose spin relaxation is faster than
the average and parts where it is slower. At medium and long times, 
the faster relaxing electron spins are already decayed, whereas the slower decaying 
electron spins remain and dominate the long-time dynamics. This effectively 
increases the decay time of the total spin polarization compared to 
the situation where only one precession frequency is present.

\begin{figure}
		\includegraphics[width=0.45\textwidth]{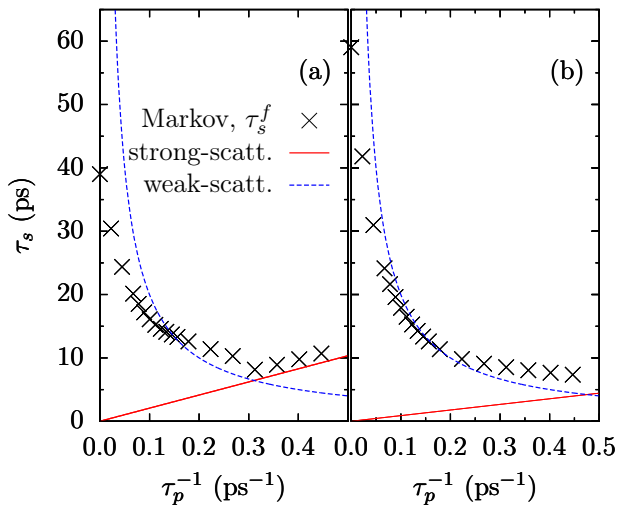}
		\caption{\label{fig:rates3} Spin decay time $\tau_s^f$ as a function of 
		the momentum scattering rate $\tau_p^{-1}$ for Gaussian initial electron 
		distributions centered at (a) $E_c=3$ meV and (b) $E_c=7$ meV above the band
		edge and with standard deviation $E_s=1$ meV.}
\end{figure}
\subsection{Influence of the central frequency of the exciting laser}
In Fig. \ref{fig:rates0}, we have studied a situation where a circularly 
polarized laser with central frequency matching the band gap was used
for the optical excitation. Now, we consider an excitation with a 
central frequency larger than the band gap and discuss the influence of
the energy difference between the laser and the band gap on the 
momentum-scattering-dependence of the spin decay time.
To this end, we repeat the above Markovian calculations of the spin decay time
$\tau_s^f$ with different values 
of the center $E_c$ of the Gaussian initial spectral electron distribution and 
extract the fitted spin decay rate $\tau_s^f$.
The results are shown in Fig. (\ref{fig:rates3}) together with the analytical
results in the strong- and weak-scattering limits.

It can be clearly seen that with increasing $E_c$ the numerical and 
analytical results agree more and more. This can be explained by the fact 
that, when the center of the peak of the electron distribution $E_c$ is increased
while its width $E_s$ remains constant, the ratio $E_s/E_c$ decreases and
the electron distribution effectively becomes spectrally sharp.

\subsection{Rashba and Dresselhaus fields}

\begin{figure}
		\includegraphics[width=0.45\textwidth]{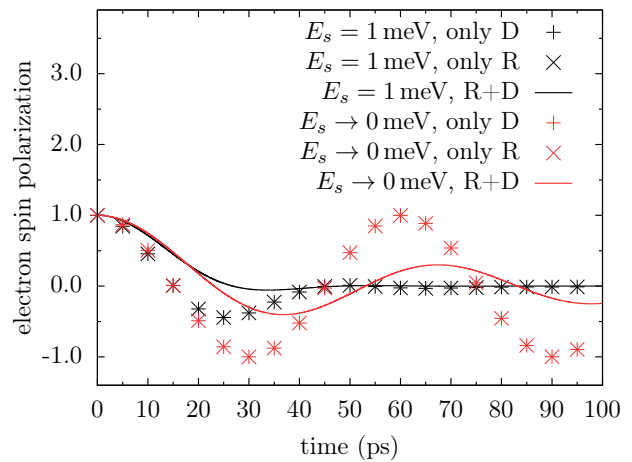}
		\caption{\label{fig:Rashba} Spin dynamics in a quantum well without 
		momentum scattering ($x=0$) subject to Dresselhaus (D), Rashba (R) or both (R+D)
		fields. The initial electron distribution is chosen to be a Gaussian 
		centered at $E_c=0$ with standard deviation $E_s=1$ meV and $E_s\to 0$,
		respectively.}
\end{figure}

The calculations presented so far only considered the Dresselhaus term 
as the origin of a $\k$-dependent effective magnetic field.
The effects of the Rashba interaction on the spin dynamics is shown in 
Fig. \ref{fig:Rashba} for an optically excited quantum well 
without momentum scattering.
It can be seen that the calculations using only the Dresselhaus field
($\alpha_R=0$, $\beta_D=1$ meVnm) and using only the Rashba interaction 
($\alpha_R=1$ meVnm, $\beta_D=0$) yield identical results. In contrast,
when both, the Rashba and the Dresselhaus terms are taken into account
($\alpha_R=\beta_D=0.5$ meVnm), the spin polarization is found to decay 
much faster. 

Even for calculations assuming a spectrally sharp initial electron distribution, the joint action of the Rashba and Dresselhaus
field results in a significant decay of the spin polarization, whereas if
the Rashba and Dresselhaus fields act alone an undamped oscillation is found. The reason for this is that, if only the Rashba or the Dresselhaus field is considered,
the magnitude of the precession frequency is fixed by the wave vector modulus $|\k|$.
When both interactions are present, this is not the case anymore and the 
magnitude of the precession frequency depends on the polar angle of the wave vector. Similar results for spectrally sharp distributions have been obtained in previous works based on rate equations\cite{Averkiev1999,Averkiev2008}.
In contrast, for distributions with finite spectral width typical in optical experiments the DIIB becomes important. As shown in Fig.~\ref{fig:Rashba}, the DIIB strongly suppresses the spin coherence for all spin-orbit fields considered here.

The fact that the impact of DIIB on the spin dynamics is similar for Rashba and Dresselhaus spin-orbit fields indicates that
the qualitative trends obtained earlier in this work for the Dresselhaus field also apply for $\k$-dependent fields of different origin.

\subsection{Algebraic decay}
\begin{figure}
		\includegraphics[width=0.45\textwidth]{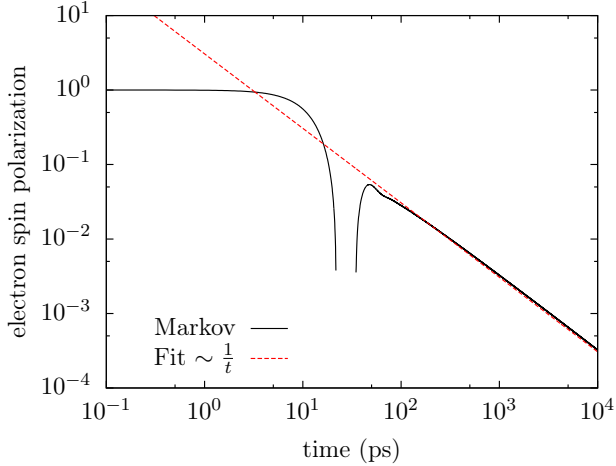}
	\caption{\label{fig:LongRange} Double-logarithmic plot of the 
	long-time behavior of the spin dynamics 
	calculated using the parameters $E_c=0$, $E_s=1$ meV, $\alpha_R=0$, $\beta_D=1$ meVnm, and $x=1.5\%$.
	}
\end{figure}

In Fig. \ref{fig:LongRange}, the spin dynamics is shown on a double-logarithmic
scale for a calculation with $x=1.5\%$ 
and $E_s=1$ meV accounting only for the Dresselhaus field.
This scale allows us to discuss the qualitative behavior of the spin dynamics on long 
timescales. 
It can be clearly seen that the spin dynamic obeys an algebraic decay 
$\propto \frac 1 t$ rather than an exponential decay at times 
$\gtrsim 100$ ps. Note that the divergence is only an artifact of negative spin polaritaztions displayed in a log-log plot.

As discussed in section \ref{subsec:Limiting-cases}, an algebraic decay is a result of
an averaging over undamped oscillatory components with a variation in the distribution
of oscillation frequencies. If these oscillations were exponentially damped individually,
a summation over the damped oscillations would also decay at least exponentially
with the smallest decay rate contained in the ensemble of damped oscillations. 
Thus, we can conclude that, on long timescales $\gtrsim 100$ ps, 
there are oscillatory components in the spin polarization that are not 
significantly damped due to momentum scattering at the impurities.

\subsection{Non-Markovian effects}
\begin{figure}
		\includegraphics[width=0.45\textwidth]{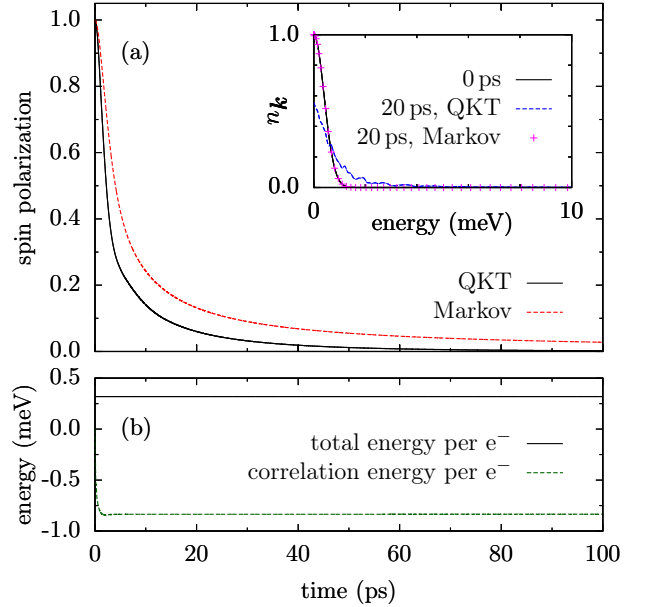}
	\caption{\label{fig:NonMarkovian} (a) Comparison of the spin dynamics according to
	the quantum kinetic theory (QKT) and the Markovian approach for $x=10\%$, 
	$d=4\,\textrm{nm}$, $\beta_D=7\,\textrm{meV}\textrm{nm}$, 
	$E_c=0\,\textrm{meV}$, and $E_s=0.4\,\textrm{meV}$ yielding a momentum
	scattering rate $\tau_p^{-1}=1.12\,\textrm{ps}^{-1}$. 
	Inset: The spectral electron distribution for the quantum kinetic and Markovian 
	calculations at $t=0$ and $t=20$ ps.
	(b) Dynamics of the average correlation energy as defined in eq.~(\ref{eq:H_corr}) and the total energy per electron
	 in the quantum kinetic theory.}
\end{figure}

The discussion of the spin decay times so far was focused on the results
of calculations based on the Markovian equations of motion (\ref{eq:DGL_Markov_sk}).
We now move on to discuss non-Markovian effects in the spin dynamics.
In Fig.~\ref{fig:rates0}(a), the momentum-scattering-dependence of
spin decay times obtained from the quantum kinetic equations (\ref{eq:DGL_CCq})
are presented together with the Markovian results. 
It is found that in a wide range of momentum scattering rates the 
Markovian and quantum kinetic calculations predict very similar spin decay 
times $\tau_s^f$. Only for large momentum scattering rates a quantitative discrepancy
is visible.

To investigate the origin of this discrepancy, the time evolution of 
the spin polarization is plotted in Fig. \ref{fig:NonMarkovian}(a) 
for a case with larger impurity concentration $x=10\%$ and therefore
large momentum scattering rates $\tau_p^{-1}= 1.12$ ps$^{-1}$.
There, the quantum kinetic result decays much faster than the Markovian
result. The reason for this is that the redistribution of carriers in $\k$-space,
which is accounted for in the quantum kinetic calculations, is completely absent in the Markovian approach. 
This redistribution can be seen in the inset of Fig. \ref{fig:NonMarkovian}(a), 
which shows the electron distribution
at $t=0$ and $t=20$ ps for both calculations.
It can be seen that the electrons are redistributed to states with on average larger 
wave vectors, which increases the average spin precession frequency and, 
in accordance with the analytical DP result (\ref{eq:tau_s_DP}), 
reduces the spin decay time.

It is noteworthy that the increase in the average wave vector implies 
an increase in the average kinetic energy, which seems at first glance to 
be at odds with the conservation of energy. However, in quantum kinetic
calculations that account for correlations, there is 
a contribution to the total energy resulting from the correlations.
\cite{FreqRenorm} Thus, the increase of the average single-particle energy is accompanied
by a corresponding build-up of negative carrier-impurity correlation
energy. This is visualized in Fig. \ref{fig:NonMarkovian}(b), where
the average correlation energy per particle, defined as
\begin{align}
\label{eq:H_corr}
\frac{1}{N_e}H_{\t{Imp}}^{\t{cor}}:=\frac{JN}{V^2N_e}\sum_{c\k\k'}\overline{C}_{c\k}^{c\k'}
\end{align}
with the total electron number $N_e=\sum_{\k \sigma} C_{\sigma\k}^{\sigma}$, is depicted as a function of time.
The total energy per electron, also shown in Fig. \ref{fig:NonMarkovian}(b),
remains constant.
It can be seen that the redistribution of carriers and therefore the
build-up of correlation energy is mostly confined to the first few picoseconds
of the dynamics. 

\section{Conclusion}

We have studied the spin dynamics in optically excited Al$_x$Ga$_{1-x}$As quantum wells
induced by the interplay of spin precession in $\k$-dependent spin-orbit fields
and momentum scattering, i.e., the D'yakonov-Perel' (DP) mechansim\cite{DP}, 
using a quantum kinetic theory.
Whereas the DP mechanism is usually only described in the strong- and weak-scattering
limits, where analytic expressions for the spin relaxation rates can be obtained,
we have investigated the dynamics over a wide range of parameters including the
limiting cases.

It is found that the time evolution of the spin polarization can be highly 
nonexponential and the notion of a decay rate for the total spin polarization
becomes ambiguous. This can be seen by the fact that two different definitions
of the spin decay time, one obtained from a fit of a stretched exponential and
one obtained from the time after which the spin polarization has decayed to
$\frac{1}{e}$ of its initial values, show quantitative and qualitative differences
in their dependence on the momentum scattering rate.

While it is common to consider only the anisotropic dependence of the spin-orbit
fields on the angle of the wave vector as a source of dephasing,
we resolve both, the angle and the modulus of the wave vector, allowing us 
to study situations with non-equilibrium carrier distributions as is the case
immediately after the optical excitation with an ultrashort laser pulse. 
This way, we also include the effects of 
dispersion-induced isotropic inhomogeneous broadening (DIIB) 
originating from the dependence of spin-orbit fields on the modulus of the
wave vector. Although DIIB has largely been ignored in the literature on the 
DP mechanism, we find that it strongly influences the spin dynamics after
ultrafast optical excitation. 

In particular, in the weak-scattering limit, where analytic expressions 
predict very large spin decay times without DIIB, the dephasing due to DIIB
limits the spin decay times even in the absence of momentum scattering. 
In the strong-scattering limit, the spin decay times are found to be longer 
than expected from the analytical result 
since the ensemble of precessing electron spins contains 
oscillatory components which decay much slower than the average electron spin and,
thus, extend the lifetime of the total spin polarization compared to 
calculations where the spectral electron distribution was assumed to be spectrally sharp and DIIB is suppressed.
Some of the oscillatory components are even found to be practically undamped and
are responsible for an algebraic decay in the long-time behavior of the total
spin polarization that cannot be measured by a spin decay time.

Whereas a linear dependence of the spin decay time on the momentum scattering
rate in the strong-scattering limit is usually considered as the hallmark of
the DP mechanism, we find that the DIIB introduces an offset leading to an
affine linear relationship between spin decay time and momentum scattering rate.
Thus, DIIB modifies central features of the DP mechansim.

Moreover, we find that DIIB can occur in situations where
the spectral electron distribution is narrow if the modulus $\Omega_\k$ 
of the $\k$-dependent precession frequency $\boldsymbol\Omega_\k$ 
depends not only on the modulus of the wave vector but also on its polar angle.
This is the case, e.g., if Rashba and Dresselhaus interaction are simultaneously
present and of comparable strength.

These findings show that DIIB  
and the effects of broad spectral electron distributions\cite{WuNingEuroPhys}, which so far are seldom discussed
in the analysis of ultrafast optical experiments dealing with DP-type spin decay,
can in fact lead to 
significant deviations from the analytical results in the strong-\cite{DP} and 
weak-scattering limits \cite{Gridnev2001}.

Although our discussion was mostly confined to the Markovian single-electron picture,
we have also presented numerical calculations taking electron-impurity correlations
explicitly into account. The non-Markovian calculations predict a faster spin decay
compared with the Markovian results. This is traced back to the build-up of 
electron-impurity correlations with negative correlation energies, which enables 
a redistribution of electrons to states with larger momentum $k$. This, in turn, 
increases the average spin precession frequency and enhances the dephasing.

In many experiments, there are other momentum scattering mechanisms to consider.
For example, phonon scattering can become important for elevated temperatures,
which gives rise to another momentum scattering channel and, in addition, also
influences the spin dynamics via the Elliot-Yafet \cite{EY1,EY2,EY_Baral} 
mechanism. Furthermore, for p-doped systems the Bir-Aronov-Pikus\cite{BAP} mechanism
affects the spin dynamics, as the electron spins interact with hole spins. 
In n-doped systems the electron spin dynamics is modified because of 
the exchange field resulting from the average carrier spins 
and the electron-electron scattering, which provides an additional 
momentum scattering mechanism \cite{SpinRelaxEE_GlazovIvchenko,Jiang2009}.

Note that most of our results are based on a Markovian description 
where the effects of momentum scattering at the impurities can be 
subsumed into a momentum scattering rate. The resulting spin dynamics does,
however, not depend on the origin of the momentum scattering. 
Thus, the same conclusions for the spin dynamics are reached when
other mechanisms are responsible for the momentum scattering, 
as long as the scattering is approximately elastic. 
For example, phonon scattering gives rise to a dissipation of energy from the
electron system and eventually leads to a thermalization of the 
electron distribution. This can reduce the average kinetic energy, the 
average wave vector, and therewith the average spin precession frequency as well as
the width of the spectral electron distribution. 
When the phonon-induced redistribution of carriers in 
$\k$-space is faster than the typical spin decay time (here: $\lesssim 50$ ps),
the electron-phonon interaction can enhance the
spin decay times since, in the strong-scattering limit, the spin decay time is 
inversely proportional to the square of the average spin precession frequency and,
in the weak-scattering limit, the spin dynamics is dominated by inhomogeneous 
broadening, which is suppressed if the width of the spectral electron distribution
is reduced. More investigations will be needed to study quantitatively the
spin dynamics in the presence of inelastic momentum scattering.

\acknowledgements
We gratefully acknowledge the financial support of the Deutsche Forschungsgemeinschaft (DFG) through Grant No. AX17/10-1.


\bibliography{bib}

\end{document}